\documentclass[a4paper,11pt]{article}
\pdfoutput=1 

\usepackage{jcappub} 
\usepackage{orcidlink}

\usepackage[T1]{fontenc} 
\usepackage{xspace}

\newcommand{\Msun}{\ensuremath{\mathit{M_\odot}}}
\newcommand{\de}{{\rm d}}
\newcommand{\Hubble}{\ensuremath{H_{0}}}

\newcommand{\fullpop}[0]{\textsc{FullPop-4.0}\xspace}
\newcommand{\Hunit}{\ensuremath{\text{km\,s}^{-1}\,\text{Mpc}^{-1}}}

\title{Heavy Black-Holes Also Matter in Standard Siren Cosmology}

\author[a]{Grégoire Pierra\orcidlink{0000-0003-3970-7970}}
\author[b]{Alexander Papadopoulos\orcidlink{0009-0006-1882-996X}}

\affiliation[a]{INFN, Sezione di Roma, 1-00185 Roma, Italy}
\affiliation[b]{Institute for Gravitational Research, University of Glasgow, \\ Glasgow, G12 8QQ, United Kingdom}

\emailAdd{gregoire.pierra@roma1.infn.it}
\emailAdd{a.papadopoulos.1@research.gla.ac.uk}

   \keywords{gravitational-waves --
                cosmology --
                standard sirens --
                mass scale --
                GWTC-4.0}

\abstract 
   {With the recent release of the largest gravitational-wave (GW) catalog to date---the Gravitational-Wave Transient Catalog (GWTC-4.0) by the LIGO-Virgo-KAGRA (LVK) collaboration---a total of 218 candidate detections of GWs from compact binary coalescences (CBCs) have been reported.
   We use 142 of these 218 events with an false alarm rate smaller than $0.25$ per year to investigate the impact of a novel parametric model to describe the full population mass spectrum of CBCs, on the estimation of the Hubble constant. This model is specially built to assess the importance of heavy black holes in GW cosmology. Our results are compared with the recent measurements reported by the LVK collaboration, both with spectral and dark sirens.
   With the spectral sirens, we estimate the Hubble constant to be $\Hubble = 81.6^{+17.6}_{-15.5}\,\Hunit$ (at $68\%$ CL), while the dark siren method yields $\Hubble=82.4^{+16.9}_{-13.3}\, \Hunit$ (at $68\%$ CL). These results represent improvements of approximately $32.9\%$ and $38.1\%$ in uncertainty, respectively. Furthermore, we demonstrate that this improvement is closely linked to the presence of an new additional mass scale for cosmology in the binary black hole (BBH) mass spectrum, found at $63.6^{+4.5}_{-4.9}\,M_{\odot}$.
   The inclusion of a new mass scale in the BH mass spectrum introduces additional constraints on the estimation of the Hubble constant, yielding precision comparable to that achieved by the two established mass scales at lower masses. Besides providing the tightest constraints on $H_0$ with standard sirens, it shows the importance of a new heavy mass feature in the BH spectrum.}

\begin{document}
\maketitle
\flushbottom

   \keywords{gravitational-waves --
                cosmology --
                standard sirens --
                mass scale --
                GWTC-4.0}
\section{Introduction}
Gravitational-wave (GW) signals from compact binary coalescence (CBCs) offer an independent method for tests of cosmological parameters, most notably the Hubble constant (\Hubble) \cite{Schutz:1986gp,Holz:2005df,Moresco:2022phi, Pierra:2025fgr}. Since September 2015 the LIGO-Virgo-KAGRA Collaboration (LVK) has detected more than 200 GW events, up to the most recent GWTC-4.0 \cite{KAGRA:2013rdx,LIGOScientific:2016dsl,LIGOScientific:2018mvr,LIGOScientific:2021usb,LIGOScientific:2025slb}. These signals provide a direct measurement of the luminosity distance ($d_{L}$) and redshifted mass of their source \cite{LIGOScientific:2025slb}. 
\par
To constrain $\Hubble$ with GWs, a redshift value for the source is required \cite{Palmese:2025zku,Pierra:2025fgr}. Several methods have been explored to obtain this, including correlation between the sky location of GW events with galaxy catalogs (the ``dark siren'' method) \cite{Schutz:1986gp,DelPozzo:2011vcw, LIGOScientific:2018gmd,Finke:2021aom, Palmese_2020, Gray_2022, Mastrogiovanni:2023emh}, or by directly extracting redshift information from an electromagnetic counterpart (the bright siren method) \cite{Holz:2005df, Dalal:2006qt,Nissanke:2009kt, LIGOScientific:2017vwq, GW170817}. The redshift can also be obtained without any external EM information, using the relationship between the detected and source frame mass spectrum, $m_\mathrm{det} = (1+z)m_\mathrm{src}$ \cite{Krolak:1987ofj, Chernoff:1993th, Mastrogiovanni:2021wsd}. This is known as the ``spectral siren'' method, providing a fully self-contained inference that requires no EM follow-up or host galaxy identification \cite{Taylor2012,LIGOScientific:2017adf,Feeney:2018mkj,Ezquiaga:2022zkx}. However it is also essential that the assumed shape of the mass spectrum mirrors the true distribution in order to avoid potential biases \cite{Mastrogiovanni:2021wsd, Pierra:2023deu,Agarwal:2024hld}. Some recent works have also explored the use of non-parametric models for the mass distribution \cite{Farah:2024xub,Tagliazucchi:2026gxn}, or for the Hubble parameter and more generally for the cosmological model, to avoid potential biases coming from the choice of the strongly parametrized model \cite{Pierra:2025hoc}.
\par
Given that accurate modeling of the CBC mass spectrum is integral to the spectral siren method, the parametric structure of our models should reflect the key features expected in the CBC population \cite{Chernoff:1993th,Taylor:2011fs,Ezquiaga:2020tns}. In particular, widely used distributions include a mass gap between neutron stars (NSs) and black holes (BHs) \cite{Farr_2011}, over-densities associated with pulsational–pair instability supernovae and stable mass-transfer channels in isolated binary evolution \cite{Callister_2024, Godfrey_2024}, and possible high-mass excesses arising from hierarchical mergers in dense environments \cite{Kimball_2021, Tiwari:2025oah}. Each of these features has the potential to improve GW-based constraints on $\Hubble$ \cite{You:2020wju,Mastrogiovanni:2021wsd,Ezquiaga:2022zkx,Pierra:2023deu,LIGOScientific:2021aug,LIGOScientific:2025jau}. According to the most recent LVK analysis, the most robust mass features are a non-empty NS–BH mass gap, an excess around $\sim10\,M_{\odot}$ and $\sim35\,M_{\odot}$, and a change in the power-law slope near $\sim35\,M_{\odot}$ \cite{Callister:2024cdx,LIGOScientific:2025pvj}. 
\par
In this paper, we analyze the GWTC-4.0 catalog using 142 confident CBC detections and a new mass model that includes an additional mass feature for heavy BHs. First, to investigate whether the data support the presence of an excess of BH at high masses, and second, to evaluate whether such a feature can be leveraged to improve spectral siren inference. Section~\ref{sec:method} presents the inference framework and our population model. In Section~\ref{sec:result}, we show how this model translates to an estimation of the Hubble constant relative to the fiducial LVK analysis of GWTC-4.0. We examine the impact of the new mass scale on the inferred astrophysical CBC population and on the standard siren measurement. Finally, in Section~\ref{sec:discussion}, we discuss the constraining power of such high mass feature with respect to the other mass scales. We conclude in Section.~\ref{sec:conclusion}.

\section{Method}
\label{sec:method}
The standard siren method enables the joint inference of the cosmological and population parameters describing the properties of CBCs, using a set of GW candidates. For all our results, we use the Python package \texttt{icarogw}, a hierarchical Bayesian pipeline designed to infer cosmological and population properties from GW data while accounting for selection effects \cite{Mastrogiovanni:2023zbw, Mastrogiovanni:2023emh}.
To account for these selection effects---that is, the finite sensitivity of the current LVK detectors---we use the publicly available injection campaign associated with GWTC-4.0. Further details on the construction and use of these injections can be found in \cite{Essick:2025zed, Pierra:2025fgr}.

\subsection{Hierarchical likelihood}
Following \cite{Mandel:2018mve, Vitale:2020aaz, Mastrogiovanni:2023zbw, Mastrogiovanni:2023emh}, the hierarchical likelihood of observing $N_{\rm obs}$ GW events, given the data set $\{\rm \boldsymbol{x}\}$, marginalizing over the local CBC merger rate $R_{0}$, can be written as
\begin{eqnarray}
    \mathcal{L}(\{\mathbf{x}\}|\mathbf{\Lambda}) \propto \prod^{N_{\rm obs}}_{i=1} \frac{\int \de \boldsymbol{\theta} \de z \mathcal{L}(\mathbf{x}_{i}|\boldsymbol{\theta},z,\boldsymbol{\Lambda})\frac{1}{1+z}\frac{\de N_{\rm cbc}(\mathbf{\Lambda})}{\de \boldsymbol{\theta}\de z \de t_{\rm s}}}{\int \de \boldsymbol{\theta}\de z p_{\rm det}(\boldsymbol{\theta},z,\boldsymbol{\Lambda})\frac{1}{1+z}\frac{\de N_{\rm cbc}(\mathbf{\Lambda})}{\de \boldsymbol{\theta}\de z \de t_{\rm s}}},
    \label{eq:likelihood_scalefree}
\end{eqnarray}
where $\mathcal{L}(\mathbf{x}_{i}|\boldsymbol{\theta},z,\boldsymbol{\Lambda})$ is the single-event likelihood, with $\boldsymbol{\theta}$ the binary’s intrinsic parameters, $z$ the redshift, and $\boldsymbol{\Lambda}$ the hyper-parameters. The denominator captures the selection effects through the probability of detection $p_{\rm det}(\boldsymbol{\theta},z,\boldsymbol{\Lambda})$ estimated with the injection campaign.
The population models describing the CBC mass distribution and CBC merger rate enter through the CBC overall rate
\begin{eqnarray}
    \frac{N_{\rm cbc}(\mathbf{\Lambda})}{\de \boldsymbol{\theta}\de z \de t_{\rm s}}=\mathcal{R}(z|\boldsymbol{\Lambda})p_{\rm pop}(\theta|z,\boldsymbol{\Lambda})\frac{1}{1+z}\frac{\de V_{\rm c}}{\de z},
\end{eqnarray}
where $\mathcal{R}(z|\boldsymbol{\Lambda})$ is the source-frame rate per comoving volume $V_{\rm c}$, chosen to be the commonly used Madau\&Dickinson like star formation rate \cite{Madau:2014bja}, and $p_{\rm pop}(\theta|z,\boldsymbol{\Lambda})$ is the population distribution, here describing solely the CBC mass distribution. This framework follows the same method used in the recent spectral and dark sirens analysis with GWTC-4.0 carried out by the LVK Collaboration \cite{LIGOScientific:2025jau}.

\subsection{Population mass model}
Astrophysically, the presence of a mass scale above $50\,M_{\odot}$ is not unfounded nor new, previous work has already proposed possible origins for it. Two main hypotheses compete: one tied to stellar evolution processes, the other to binary black hole formation channels. Supported by population synthesis codes, \cite{Torniamenti:2024uxl} predicts an mass excess at $70\,M_{\odot}$ coming from 2nd-generation hierarchical mergers and consistent with \cite{Antonini:2022vib, Li:2025iux, Plunkett:2026pxt}.
In \cite{Antonini:2025zzw}, this mass excess is also presented as the sign of a mass gap associated with pulsational pair-instability supernovae (PPISN): stellar evolution theory predicts no BH formation between roughly $40-130\,M_{\odot}$ from single massive star collapse, yet event are observed in this range, pointing to a new formation channel above $50\,M_{\odot}$ that less flexible models were not testing.
This new model allows us to assess whether such an accumulation of BHs at high masses is supported by the data using a parametric model, and whether this feature influences the estimation of cosmological parameters such as the Hubble constant.

This population model builds on the \fullpop from \cite{LIGOScientific:2025jau}, which in turn is an extension of the \textsc{PowerLaw--Dip--Break} from \cite{Fishbach_2020} and \cite{Farah_2022}. Like its predecessors, it covers the whole CBC mass range (NSs+BHs) \cite{Mali:2024wpq} and employs two identical distributions for the primary and secondary masses, combined with a pairing function which ensure the usual condition $m_1>m_2$ \cite{Fishbach:2019bbm}. Its basic brick is a broken power-law with two Gaussian peaks, and we develop this by increasing the number of Gaussian peaks to three. The first power-law models the low-mass region for NS-containing events, coupled to a second power-law for high-mass events containing BHs, and the two are connected via a dip function playing the role of a potential mass gap between NSs and BHs. 
A detailed description of the functional forms of the basic building blocks of the mass model, like the parameterization of the dip between the NSs and BHs, can be found in Appendix C of \cite{LIGOScientific:2025jau}. 
In total, this model is characterized by twenty-two population parameters, including parameters governing the dip junction and the low-mass smoothing. 
Therefore, the mass distribution can be expressed as
\begin{equation}
\label{eq:massdist}
\begin{split}
p_{\rm pop}(m_{1,\mathrm{s}}|\mathbf{\Lambda}) &=
(1-\lambda_{\rm g})\mathcal{B}(m_{1,\mathrm{s}}|m_{\rm min},m_{\rm max},\alpha_1,\alpha_2,b,\boldsymbol{\Lambda}) \\ &+ \lambda_{\rm g}\lambda_{1}
\mathcal{G}_{1}(m_{1,\mathrm{s}}|\mu^{1}_{\rm g},\sigma^{1}_{\rm g}) \\
& + \lambda_{\rm g}(1-\lambda_1)\lambda_2
\mathcal{G}_{2}(m_{1,\mathrm{s}}|\mu^{2}_{\rm g},\sigma^{2}_{\rm g}) \\
& +\underbrace{\lambda_{\rm g}(1-\lambda_1)(1-\lambda_2)
\mathcal{G}_{3}(m_{1,\mathrm{s}}|\mu^{3}_{\rm g},\sigma^{3}_{\rm g})}_\text{Additional term relative to \fullpop}\,, \\
\\ 
p_{\rm pop}(m_{2,\mathrm{s}}|\mathbf{\Lambda})&=p_{\rm pop}(m_{1,\mathrm{s}}|\mathbf{\Lambda}).
\end{split}
\end{equation}
Where $\mathcal{B}$ is a broken power-law, $\mathcal{G}_{i\in[1,2,3]}$ are the Gaussian components, and the weights between the Gaussian feature and the rest of the distribution are given by $\lambda_{i\in[1,2]}$. In particular, we use the notation $(\mu_{\rm g}^{i},\sigma_{\rm g}^{i})$ to refer to the mean and variance of each Gaussian component $\mathcal{G}_i$. The overall distribution also include a low and high mass tapering to avoid a non-physical sharp cut-off.
The extra population parameters governing the dip and the low and high mass tapering, not explicitly shown in Eq.~\ref{eq:massdist}, are contained in the extra $\boldsymbol{\Lambda}$ of the broken power-law to simplify the notation.
\par
To ensure that all Gaussian peaks receive equal prior weight, we place a Dirichlet prior on the mixture weights defined in Eq.~\ref{eq:massdist}. In this framework, we assign a Beta prior ($\alpha=1,\beta=2$) to $\lambda_{1}$ and a uniform prior to $\lambda_{\rm g}$ and $\lambda_2$, so that the combined peak weights correspond to the elements of a symmetric Dirichlet distribution \cite{Kotz:2000dir}.
   
\section{Results\label{sec:result}}
Our results are produced using 142 CBCs from the GWTC-4.0 catalog, selected using a FAR$<0.25$yrs \cite{LIGOScientific:2025slb}. Even though this event selection excludes the GW event GW231123, in line with \cite{LIGOScientific:2025slb}, we still perform an extra analysis including it. We compare the results obtained with our new population model with the LVK multi-population model, looking at both spectral and dark siren inferences \cite{LIGOScientific:2025jau}. For the dark siren approach, we consider the K-band of the GLADE+ galaxy catalog, with non-uniform luminosity weights \cite{Dalya:2018cnd, Dalya:2021ewn}, which is identical to the fiducial galaxy catalog used in \cite{LIGOScientific:2025slb}. To estimate the selection effect, we use the publicly available injection campaign from LVK associated to GWTC-4.0 \cite{LIGOScientific:2025yae, Essick:2025zed}. All our results and numerical values are stated as the maximum a posteriori (MAP) value along with the $68\%$ credible level (C.L.) unless otherwise stated.

\begin{figure}[ht!]
\centering
  \includegraphics[width=0.9\textwidth]{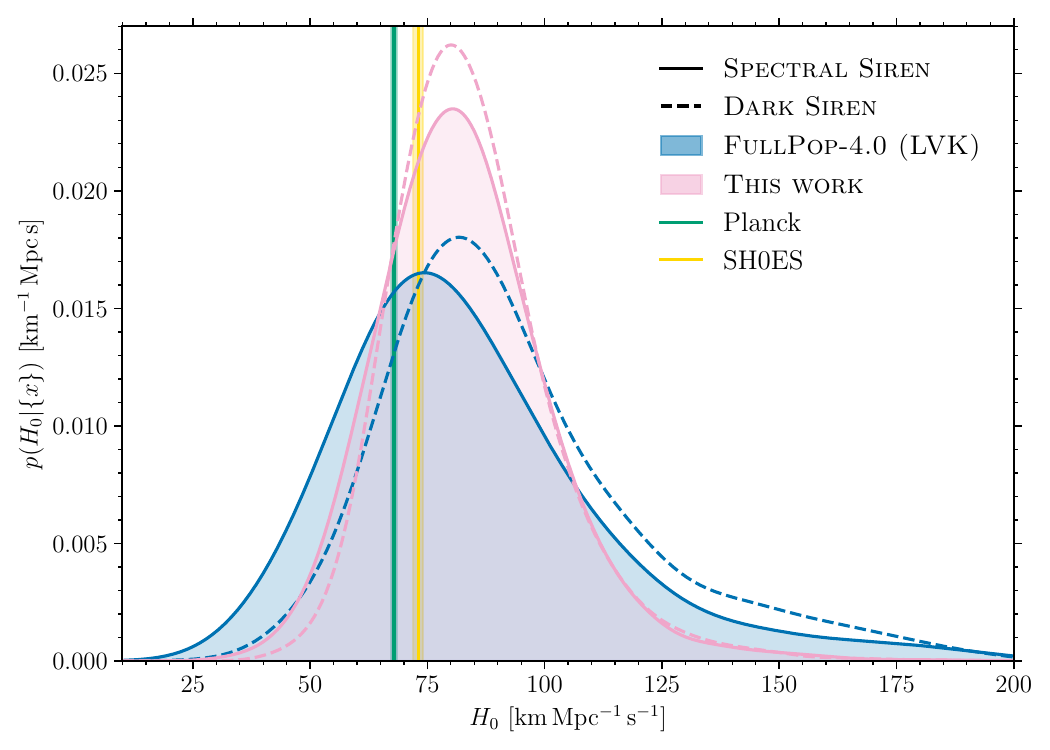}
  \caption{Marginalized Hubble constant posteriors for different analyses. The fiducial GWTC-4.0 model with $\fullpop$ mass model is shown in blue, and this work shown in pink. The solid lines with filled curves present the spectral siren posteriors, with the dark siren posteriors presented with the dashed lines. Vertical lines are the Hubble tension reference values from Planck and SH0ES \cite{Planck:2015fie, Riess:2021jrx}.}
     \label{Fig:H0 posterior}
\end{figure}

\subsection{The Hubble constant\label{sec:hubbleresult}}
Figure~\ref{Fig:H0 posterior} shows the marginalized posteriors of the Hubble constant obtained with spectral and dark sirens, using our new mass model and comparing to LVK's multi-population model. We also indicate the recent reference values defining the Hubble tension from Planck and SH0ES \cite{Planck:2015fie, Riess:2021jrx}. Analyses with our new mass model yield a posterior of $\Hubble=81.6^{+17.6}_{-15.5}\,\Hunit$ in the spectral siren case, improving to $\Hubble=82.4^{+16.9}_{-13.3}\,\Hunit$ in the dark siren case. This corresponds to a $32.9\%$ ($38.1\%$) improvement in the Hubble constant constraint for the spectral (dark) siren analysis. This notable improvement is closely linked to the reconstruction of a new mass scale at high masses, which we discuss in section.~\ref{sec:massscales}. 
Combining the constraints from our model with the bright siren GW170817 gives $\Hubble=76.7^{+12.5}_{-9.6}\,\Hunit$ ($\Hubble=77.1^{+12.0}_{-9.2}\,\Hunit$) for the spectral (dark) siren case, corresponding to an $7.4\%$ ($14.5\%$) improvement relative to LVK's latest result. 
\par
While this parametric model improves the GWTC-4.0 measurement of the Hubble constant, it is still insufficient to resolve the Hubble tension.

\subsection{The heavy black-hole mass scale}\label{sec:massscales}
As shown in the previous section, introducing an additional mass scale into the BH mass spectrum leads to a tighter constraint on $\Hubble$. We now turn to quantifying the extent of this improvement. 
\par
\begin{figure}[ht!]
    \centering
    \includegraphics[width=\textwidth]{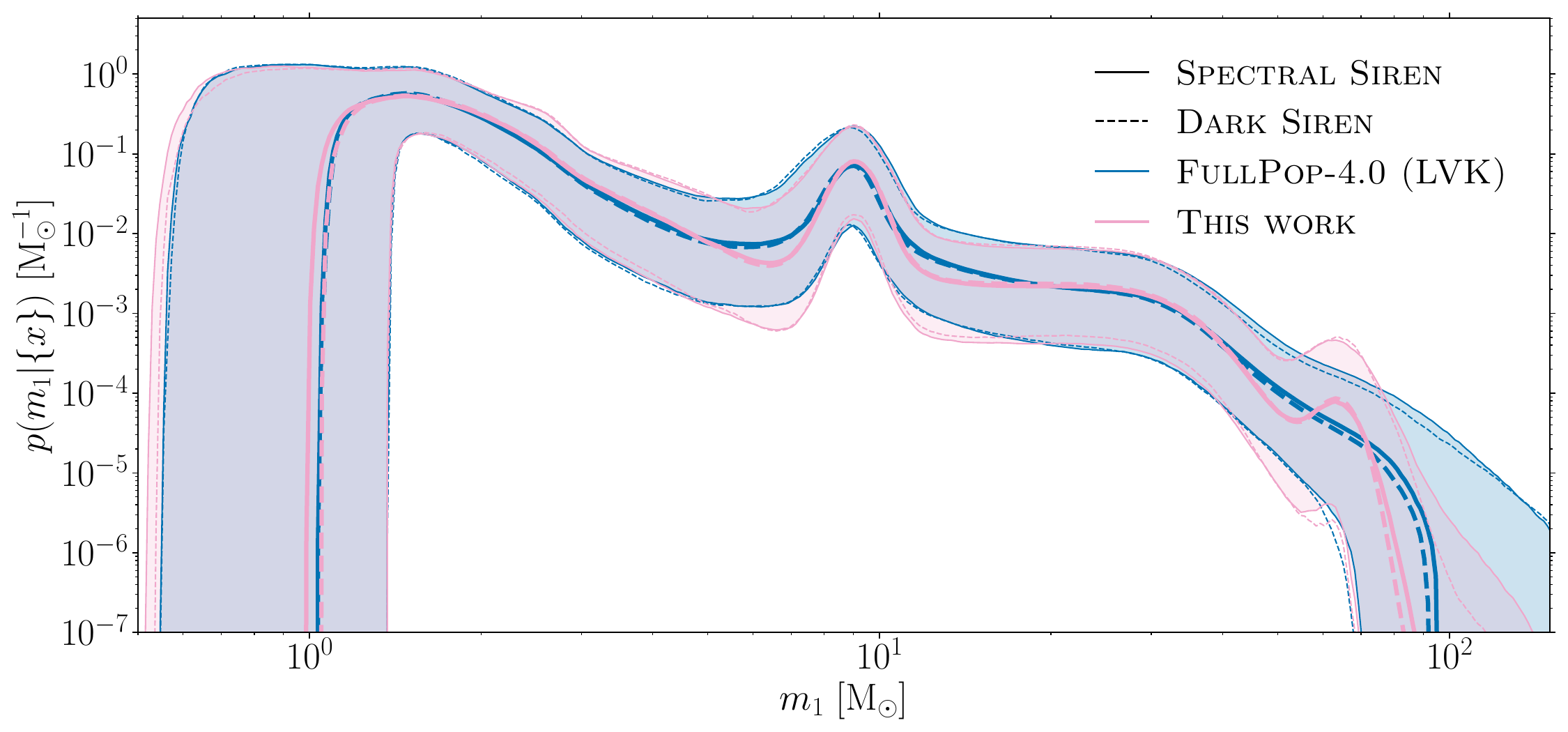}
    \caption{Posterior predictive distribution of the primary mass spectrum, with the $\fullpop$ in blue and our model in pink. The thick lines denote the median shape of the distribution, while the filled regions give the $68\%$ C.L.}
    \label{Fig:PPC}
\end{figure}

Fig.~\ref{Fig:PPC} shows a comparison of the reconstructed mass distributions from the spectral and dark siren inferences on GWTC-4.0. All the mass spectra are in very good agreement, particularly in the NS region and the low-mass BH region (below $50 M_{\odot}$). However, in the high-mass region of the distribution, we observe an extra accumulation of BHs around $63 M_{\odot}$, clearly visible in both the spectral and dark siren analyses.
Our model is designed to be flexible: it does not force the presence of a mass feature and allows the data to determine whether a peak is supported. To ensure an agnostic parametric inference of possible mass scales, the prior ranges for the means and variances of all Gaussian peaks span the entire BH mass spectrum, from $5\,M_{\odot}$ to $150\,M_{\odot}$.

\begin{figure}
\centering
  \includegraphics[width=0.9\textwidth]{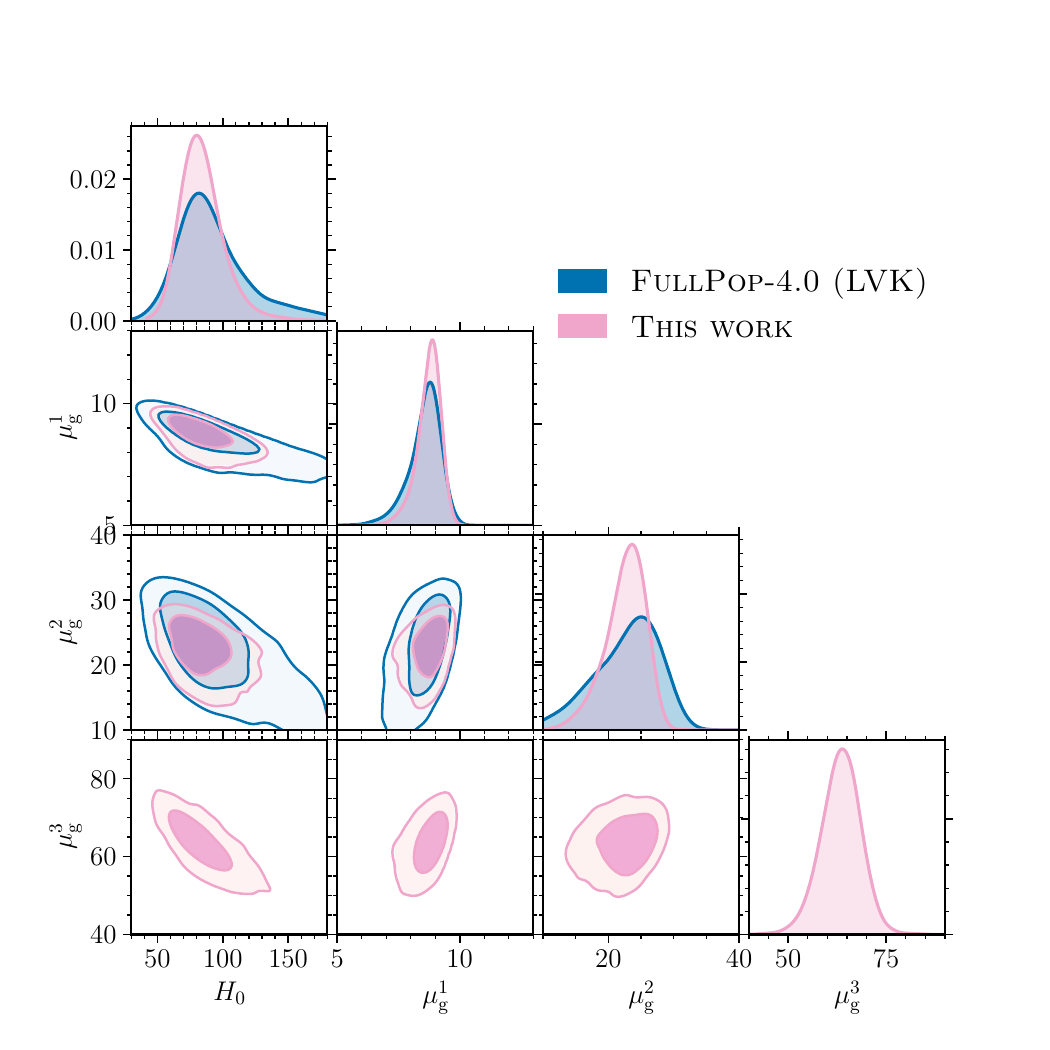}
  \caption{Inferred 1D posteriors and 2D-contours of the Hubble constant $H_0$ and the position of the different mass scales, for both the $\fullpop$ (blue) and our mass model (pink). The results were obtained with the dark siren inference. The contours shows the $1\sigma$ and $2\sigma$ levels.}
     \label{Fig:Corner}
\end{figure}

The impact of this high-mass scale is illustrated in Fig.~\ref{Fig:Corner}, which shows the correlations between each mass feature and the Hubble constant obtained from the dark siren inference. The spectral siren results are not shown on this figure, but are fully consistent with the dark siren ones. From Fig.~\ref{Fig:Corner}, we first observe that the additional mass feature is located exactly at $63.6^{+4.5}_{-4.9}\,M_{\odot}$ (not overlapping with the other two features), and it exhibits a strong anti-correlation with $\Hubble$. This anti-correlation behavior was anticipated, as previous studies have highlighted the role of mass scales in standard siren inference \cite{LIGOScientific:2021aug, Pierra:2023deu,LIGOScientific:2025jau}. 
Secondly, comparing with the $\fullpop$ model, we see a tightening of the constraints on the means of the other two mass features at lower masses, while their locations remain consistent. Specifically, we find the first mass scale at $8.8^{+0.3}_{-0.5}\,M_{\odot}$ and the second at $22.9^{+2.5}_{-3.4}\,M_{\odot}$. Although the first two mass scales are better constrained, they still show a strong correlation with the Hubble constant. In other words, the reconstruction of a new mass feature at high masses also better constrains the inferred mass spectrum at lower masses, hence also improving the Hubble constant estimation.
We note that in terms of relative weights inferred in each Gaussian peak, we find that the probability in the lower and middle mass peaks dominated compared to that in the high-mass peak, with less than $4\%$ of the total probability contained in the $63\,M_{\odot}$ component.
\begin{table}
\centering
\renewcommand{\arraystretch}{1.25}

\begin{tabular}{p{0.3\linewidth}ll}
\hline
\textbf{Model} & \textbf{$A(\Hubble, \mu_{\rm g}^{i})$} & \textbf{$r(\Hubble, \mu_{\rm g}^{i})$} \\ \hline\hline
\fullpop (LVK) & 
\begin{tabular}[t]{@{}l@{}}
$\mu_{\rm g}^1$: 100.2 $\pm$ 15.0 \\ 
$\mu_{\rm g}^2$: 615.9 $\pm$ 9.8 \\ 
$\mu_{\rm g}^3$: -- 
\end{tabular}
&
\begin{tabular}[t]{@{}l@{}}
$\mu_{\rm g}^1$: 0.162 $\pm$ 0.023 \\ 
$\mu_{\rm g}^2$: 0.138 $\pm$ 0.002 \\ 
$\mu_{\rm g}^3$: -- 
\end{tabular}
\\ \hline
\text{This work} &
\begin{tabular}[t]{@{}l@{}}
$\mu_{\rm g}^1$: 58.1 $\pm$ 10.6 \\ 
$\mu_{\rm g}^2$: 291.4 $\pm$ 6.7 \\ 
$\mu_{\rm g}^3$: 393.1 $\pm$ 19.9
\end{tabular}
&
\begin{tabular}[t]{@{}l@{}}
$\mu_{\rm g}^1$: 0.196 $\pm$ 0.028 \\ 
$\mu_{\rm g}^2$: 0.736 $\pm$ 0.029 \\ 
$\mu_{\rm g}^3$: 0.213 $\pm$ 0.008
\end{tabular}
\\ \hline
\end{tabular}
\caption[]{Table of metrics for comparing the effect of peaks on $\Hubble$ estimation, including the  $1\sigma$ contour of the 2D posterior areas in $[\Hunit\,\Msun]$ (first column) and axis ratios (second column). These assume the posteriors are approximately gaussian, and include $1\sigma$ numerical errors.}
\label{tab:mi_areas}

\end{table}

We verify this effect by quantifying the correlations with $\Hubble$, using the metrics shown in Table~\ref{tab:mi_areas}, based on the 2D-posterior areas and axis-ratios computed from the covariance matrix (assuming approximately Gaussian posteriors). The 2D area reflects how tightly each mass scale is constrained with $\Hubble$, while the axis ratio serves as a proxy for the strength of their correlation. We chose to quantify correlations using a combination of the 2D areas and the axis ratio, because commonly used proxies such as the Pearson coefficient can assign the same value to posteriors with markedly different 2D areas, and therefore do not reflect the overall tightening of the constraints \cite{benesty2009pearson}.
With the inclusion of a third mass scale, both $\mu_{\rm g}^1$ and $\mu_{\rm g}^2$ show a substantial reduction in area, from $100.2$ to $58.1$ and from $615.9$ to $291.4$ (in $\Hunit\,\Msun$), respectively. The axis ratios reveal that $\mu_{\rm g}^1$ weakens its degeneracy with $\Hubble$ (from $0.162$ to $0.196$), the same is true of $\mu_{\rm g}^2$ - changing from $0.138$ to $0.736$. 
The new mass scale exhibits a larger posterior area, $393.1\,\Hunit\,\Msun$, and an axis ratio of $0.213$, indicating an additional strong degeneracy with $\Hubble$. However, its main impact is to redistribute the correlations: it reduces the areas of $A(\Hubble, \mu_{\rm g}^1)$ and $A(\Hubble, \mu_{\rm g}^2)$, while creating a third mass scale. Lastly, we find that our constraint of the maximum BH mass is weakened by the additional peak, whilst still being in agreement with the \fullpop constraint. 
\par
Finally, the Bayes factors show that our new population model is mildly preferred over the $\fullpop$ model used by the LVK, despite its higher dimensionality. As discussed in \cite{LIGOScientific:2025jau}, by applying narrower priors to both models yields a Bayes factor of $\mathcal{B}=6$ in favour of the model with an additional mass scale once reweighted by the effective prior volume to ensure its correctness. This can be interpreted as "substantial" support from the data in favour of an additional peak, according to \cite{Kass1999}. Such mass scale has also been reported in other works, for example in \cite{MaganaHernandez:2024qkz} with a smaller data set, they find support for a feature around $65-70\,M_{\odot}$, preferred by the Bayes factor and supported by a simulation study. In \cite{LIGOScientific:2025pvj}, using both strongly and weakly modelled inference, they report evidence for a feature around $60\,M_{\odot}$

\section{Discussion}\label{sec:discussion}
\begin{figure*}[ht!]
\centering
  \includegraphics[width=\textwidth]{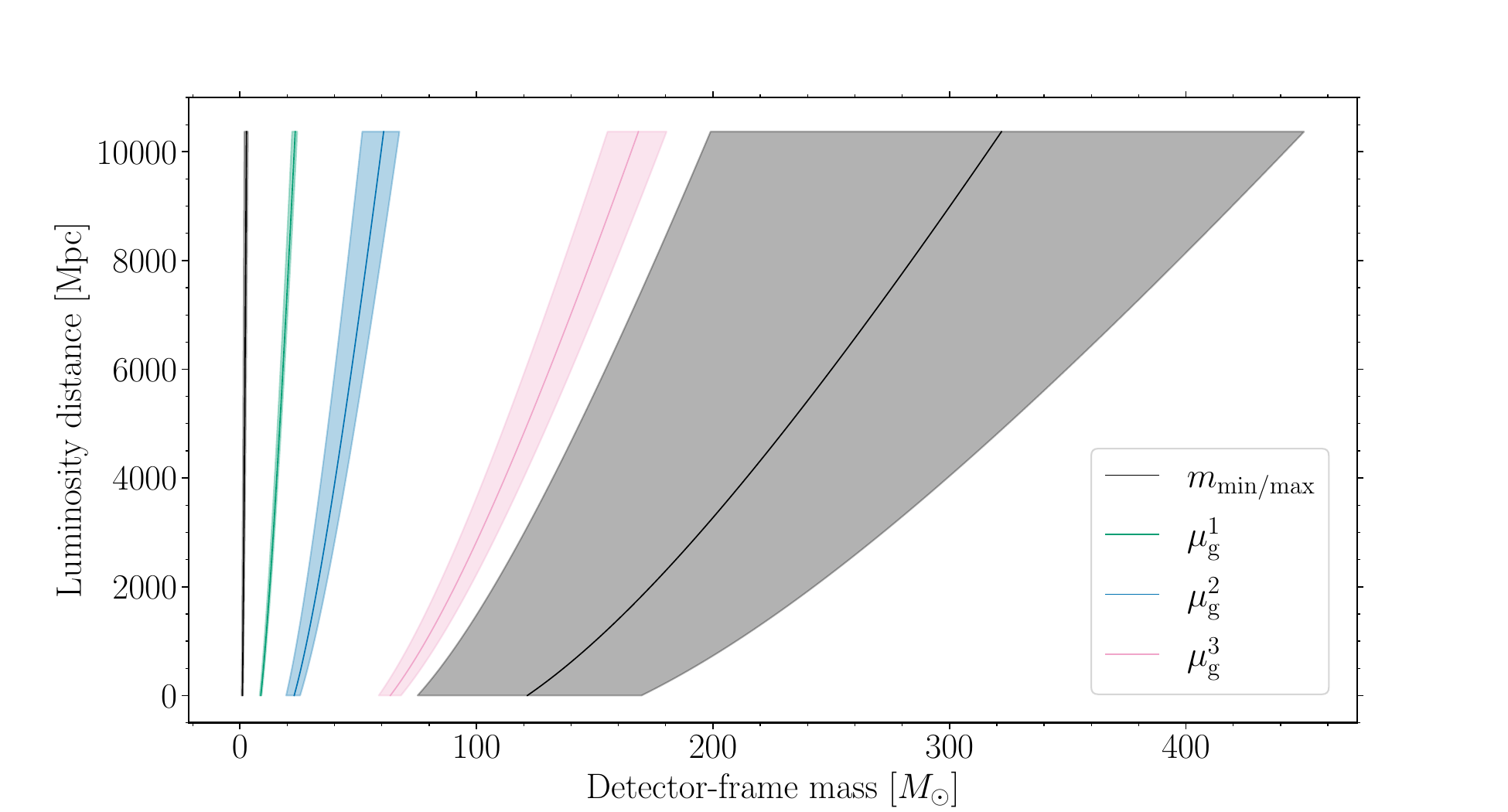}
  \caption{Mass scale evolution estimated from the dark siren inference of GWTC-4.0, visualized in the detector frame $\{d_{\rm L}, m_{\rm det}\}$. The solid lines represent the MAP value of the different mass scales of our population model, namely the minimum and maximum masses (black), the first (green), second (blue) and third (pink) Gaussian peaks. The shaded regions show the $68\%$ C.L.}
  \label{Fig:MassScale}
\end{figure*}

Quantifying the contribution of each mass scale to the overall estimation of the Hubble constant is a challenging task. Recent studies have begun to explore this question. For instance, \cite{Mancarella:2025uat} investigates the impact of individual GW events on $\Hubble$, while \cite{Mali:2024wpq} examines in detail the correlation between the $30\,M_{\odot}$ feature and $\Hubble$.
Here, rather than focusing on individual events or on deriving correlation coefficients, we aim to understand how the different inferred mass scales contribute to the resulting cosmological constraints. In \cite{Pierra:2025hoc}, it is explained why the constraining power of the standard siren method arises from the evolution of the mass spectrum (and in particular mass features) when looked in the detector frame, tracked jointly in the $\{d_{\rm L}, m_{\rm det}\}$ parameter space. Mass scales that show strong evolution across this space contribute more significantly to the final constraints on the cosmological parameters. Conversely, if the mass spectrum were constant across all distances, the method would provide little information. Another important factor is how well each scale is constrained, so the most informative mass scales are those that both exhibit strong evolution in the detector frame and are tightly inferred from the GW data.
\par
Fig.~\ref{Fig:MassScale} shows the inferred evolution of the main mass features in the $\{d_{\rm L}, m_{\rm det}\}$ parameter space, namely the minimum and maximum masses, and the three Gaussian components of our new population model. This evolution is reconstructed from posterior samples of the mass distribution parameters and $\Hubble$. For each sample, the inferred source-frame mass scales are converted to detector-frame using $m_{\rm det} = (1+z)\, m_{\rm src}$, while the corresponding luminosity distance is computed from the redshift, using the $\Hubble$ posterior. This approach allows us to visualize the evolution of the mass scales and their uncertainties in the parameter space relevant to the detected signals.
Based on Fig.~\ref{Fig:MassScale}, we can draw the following conclusions: (i) the minimum mass shows no evolution in the $\{d_{\rm L}, m_{\rm det}\}$ space, which explains why this mass scale does not usually correlate with $\Hubble$, therefore does not contribute to its constraining power. In contrast, the maximum mass evolves strongly, but its large uncertainties limit its impact, consistent with the typical $H_0$–$m_{\rm max}$ correlation observed with standard siren \cite{LIGOScientific:2025jau}. (ii) All three Gaussian components exhibit some evolution with $d_{\rm L}$, with the third peak displaying the strongest. (iii) The first peak, while evolving slightly less than the others, has the smallest uncertainties, hence has equivalent constraining power as the third one. 
These conclusions are consistent with the metrics derived in Section~\ref{sec:result} and demonstrate that the constraining power of a mass scale is governed by the interplay between its evolution in the $\{d_{\rm L}, m_{\rm det}\}$ space and how well it can be inferred from the GW data.
\par
As an additional test, we also investigated the inclusion of the special GW candidate GW231123 \cite{LIGOScientific:2025rsn}. This event, with a total detector-frame mass exceeding $200\,M_{\odot}$, was excluded in the recent LVK cosmology analysis. Since our study focuses on the role of heavier BHs, GW231123 could potentially have an impact here. Re-running all our analyses including this event, we find that the improvement in the Hubble constant constraints decreases to $8.3\%$ and $4.1\%$ for the dark and spectral approaches, respectively. In particular, we observe that the inclusion of this massive event pushes the inferred maximum mass to $\sim170\,M_{\odot}$, and that the second and third Gaussian peaks become less well constrained compared to the analysis without this event. This may indicate that even our extended mass model struggles to accommodate such an extreme system within the overall population.
\par
We verified that our results are numerically stable with respect to the evaluation and convergence of the hierarchical Bayesian framework. In particular, we reproduced our analyses using a cut on the variance of the log-likelihood, such that $\mathrm{var}(\mathrm{log} \mathcal{L}) < 1$, which was shown as a reliable threshold in \cite{Talbot:2023pex}. The details of these numerical stability test are explained in Appendix.~\ref{app:numstabi} We find that our results remain in perfect agreement when this threshold is applied as shown in Fig.~\ref{Fig:H0 posterior nume stab}.
\par
Besides, our inferred mass spectrum does not exhibit the pronounced feature reported in \cite{MaganaHernandez:2025cnu}, where their semi-parametric approach indicates an excess around $\sim75\,M_{\odot}$ followed by a sharp drop before $80\,M_{\odot}$. Such feature is in disagreement with our results, and it also appears to be in tension with their parametric model as well as with other recent studies using both parametric and non-parametric mass models \cite{LIGOScientific:2025jau,LIGOScientific:2025pvj, Tagliazucchi:2026gxn, Tiwari:2025oah} with GWTC-4.0 data. This discrepancy, seen in both our results and other independent analyses, suggests that the reported excess in \cite{MaganaHernandez:2025cnu} is likely model-induced and therefore their constraints on the Hubble constant are over-estimated.
\par
We would also note that while this manuscript was in preparation, the model introduced here was included in the GWTC5-5.0 cosmological analyses \cite{LIGOScientific:2026gwtc5}, using an expanded dataset of 235 GW events, almost 100 more than utilised in this work. This analysis, using the same pipeline and setup as we present, but with a larger dataset, also found that the model with an additional peak is favoured by the data and produces more stringent constraints on the Hubble constant when compare to the \fullpop. Additionally, the existence of this feature in GWTC-5.0 is further corroborated with the non-parametric method presented in \cite{Leyde:2026cosmopyro}. These results further strengthen the evidence for the existence of a heavy BH mass scale in the CBC mass spectrum.
\par
Finally, even though the scope of this work is linked to how a high-mass feature can help constrain the Hubble constant, we note that the support we find for a BH excess at $63\,M_{\odot}$ might be the signature of a sub-population of BHs originating from previously merged BHs, also known as hierarchical mergers, in agreement with recent studies on GWTC-4.0 \cite{Li:2025iux, Plunkett:2026pxt} as well as earlier predictions for the signature of hierarchical mergers \cite{Gerosa:2021mno}.

\section{Conclusions \label{sec:conclusion}}
In this work, we have demonstrated that heavy BHs have a strong impact on standard siren cosmology, for both spectral and dark sirens. Specifically, we have shown that the BH mass spectrum exhibits an excess around $63\,M_{\odot}$, and that this feature can be used as a new mass scale in GW cosmology when an adequate population model is employed. 
We find that the constraint on $H_{0}$ improves by $38.1\,\%$ with 141 dark sirens, relative to the latest LVK result using the GWTC-4.0 catalog. When combined with the bright siren GW170817, the improvement reaches $14.5\,\%$, yielding $H_{0}=77.1^{+12.0}_{-9.2},\Hunit$. Moreover, we show that the additional cosmological constraining power provided by our model is not solely due to the new mass scale at $63.6^{+4.5}_{-4.9}\,M_{\odot}$, but also to the improved constraints on the two other mass scales around $\sim 9\,M_{\odot}$ and $\sim 23\,M_{\odot}$ triggered by the more precise reconstruction of the high mass region. 
Finally, we quantify the relative importance of each mass scale for constraining the Hubble constant. In particular, we show that, when examined in the detector-frame plane, the constraining power of a mass scale is determined by the combination of its evolution in that plane and how well it can be measured, making the $63\,M_{\odot}$ scale a newly important one. Future standard-siren analyses should therefore consider models capable of capturing this feature.

\acknowledgments
G. Pierra is supported by ERC grant GravitySirens 101163912. Funded by the European Union.
Views and opinions expressed are however those of the author(s) only and do not necessarily reflect
those of the European Union or the European Research Council Executive Agency. Neither the
European Union nor the granting authority can be held responsible for them. A. Papadopoulos is supported by UKRI STFC studentship 323353-01. A. Papadopoulos thanks Rachel Gray and Chris Messenger for their useful discussions. This material is
based upon work supported by NSF’s LIGO Laboratory which is a major facility fully funded by the National Science Foundation. The authors are grateful for computational resources provided by the LIGO Laboratory and supported by National Science Foundation Grants PHY-0757058 and PHY-0823459.

\begin{appendix}
\section{Numerical stability}
\label{app:numstabi}
In this appendix, we provide additional details on the different numerical stability tests related to the likelihood evaluation, and therefore on the overall robustness of our hierarchical inferences. In \texttt{icarogw}, the integrals appearing in the likelihood in Eq.~\ref{eq:likelihood_scalefree} are estimated numerically by summing over single-event PE samples via Monte-Carlo (MC) integration \cite{Mastrogiovanni:2023zbw}. To ensure the accuracy of this approximation, one can impose either a threshold on the effective sample size $\rm N_{\rm eff}$ \cite{Farr:2019rap,Essick:2022ojx}, or alternatively a threshold on the total variance of the population log-likelihood $\rm var(ln\,\mathcal{L})$, which has been shown to provide an even more conservative stability criterion \cite{Talbot:2023pex,Heinzel:2025ogf}.
\par
Following standard prescriptions for GW population studies, for the main results of this paper we require a minimum of $\rm N_{\rm eff}>10$ effective PEs and $\rm N_{\rm eff}>4N_{\rm obs}$ effective injections. While when using the variance-based criterion, we impose $\rm var(ln\,\mathcal{L})<1$.
\begin{figure}[ht!]
\centering
    \includegraphics[width=0.9\textwidth]{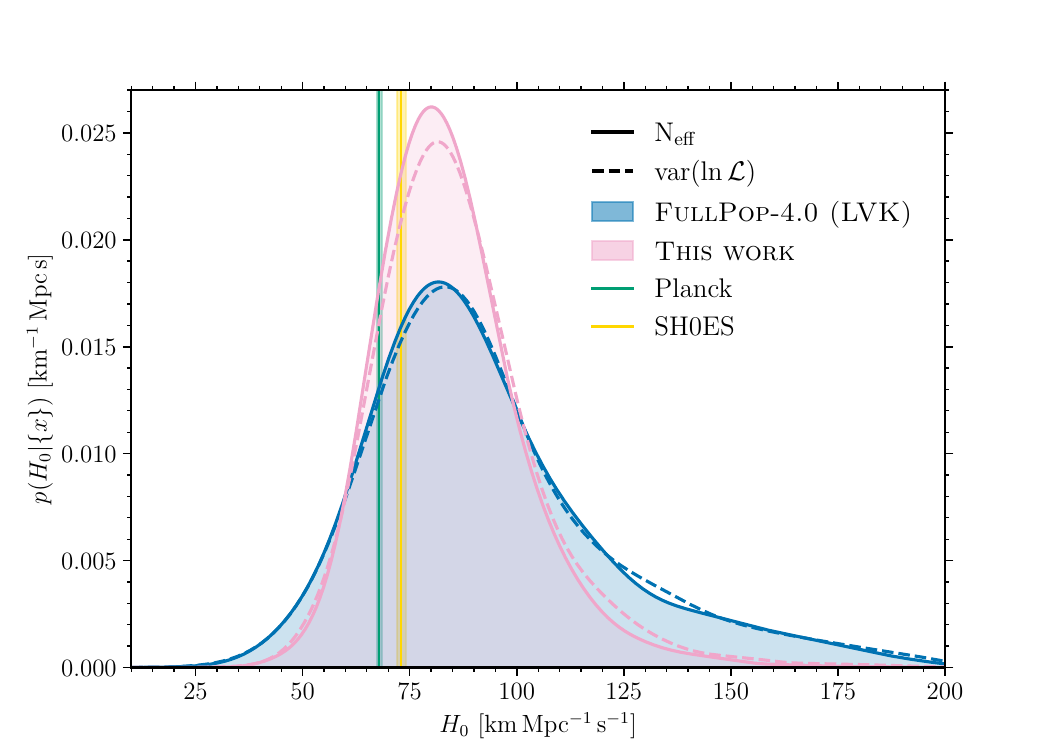}
    \caption{Marginalized Hubble constant posteriors obtained via dark siren inference, for different numerical stability tests. The fiducial GWTC-4.0 model with $\fullpop$ mass model is shown in blue, and this work shown in pink. The solid lines are derived using a cut on the number of effective PE samples, while the dashed posteriors using a cut on the variance of the log-likelihood. Vertical lines are the Hubble tension reference values from Planck and SH0ES \cite{Planck:2015fie, Riess:2021jrx}.}
    \label{Fig:H0 posterior nume stab}
\end{figure}
Figure.~\ref{Fig:H0 posterior nume stab} shows the inferred marginalized posteriors of the Hubble constant obtained with both the \textsc{FullPop-4.0} and our new mass models, using alternatively the numerical stability cuts based on the effective sample size and on the variance of the log-likelihood for dark sirens with the GLADE+ galaxy catalog. The agreement between the $H_0$ posteriors (as well as all other population parameters) obtained with the two criteria demonstrates that the MC integration, and therefore the likelihood integrals, are numerically stable and lead to consistent inference results for both models.
\end{appendix}


\end{document}